\documentclass[twocolumn,showpacs,aps,prl]{revtex4}
\usepackage{amssymb}
\usepackage{graphicx}
\usepackage{dcolumn}
\usepackage{bm}
\usepackage{color}

\begin{document}

\title{Magneto-thermoelectric current induced by phonon drag in low-dimensional
junctions}
\author{S.~E.~Shafranjuk}
\affiliation{Department of Physics and Astronomy, Northwestern University, Evanston, IL 60208}

\date{\today }


\begin{abstract}
We examine splitting the heat flow in a low-dimensional junction under influence an external d.c. magnetic field. The junction is a crossing between the narrow single atomic layer stripe (or a nanotube) of a semiconductor C with a metal stripe N (C/N-knot). External source of heat injects the non-equilibrium (NE) phonons, electrons, and holes into C which then propagate in direction the C/N-knot. Most of the heat is carried by NE phonons which drag additional electron and hole excitations along C. In vicinity the C/N-knot, the Lorentz force pulls the charge carriers from C to N thus inducing a substantial lateral magneto-thermoelectric electric current (MTEC) along N.
\end{abstract}

\pacs{84.60.Rb, 73.40.Gk, 73.63.Kv, 44.20.+b}
\maketitle

Magneto-thermoelectric phenomena attract significant attention of many researchers because they provide fundamental knowledge about transport the charge carriers and phonons in low-dimensional conductors.~\cite{Kane-Fisher,Kane-Mele} Thermoelectricity also finds a variety of applications in scientic research and engineering. Controlling the heat flow by an external field is important, e.g., in processes of thermoelectric cooling and energy generation.~\cite{TEG-book,P-Kim,Dressel,Shafr-TEG,Shafr-Review}
\begin{figure}
\includegraphics[width=75 mm]{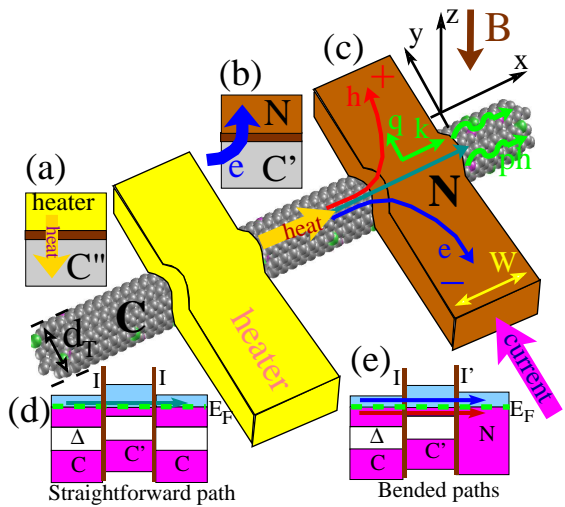}
\caption{{Color online. (a)~Heat injection into the C$^{\prime \prime}$-section under metal N. (b)~Electron tunneling from the C$^{\prime}$-section into the metal stripe  N.  (c)~Main figure: Splitting the heat flow by the C/N-junction (C/N-knot in the nanotube-stripe cross geometry) when a finite d.c. magnetic field $\mathbf{B}$ is applied in perpendicular to the knot's plane. (d) and (e) are two different energy diagrams wich correspond to distinct electron trajectories C/C$^{\prime}$/C and C/C$^{\prime}$/N correspondly.}} 
\label{fig1} 
\end{figure}
On this path, novel single atomic layer materials have a great potential.~\cite{P-Kim,Dressel,Shafr-TEG,Shafr-Review} 

In this Letter we show that the heat stream along a narrow single atomic layer semiconducting stripe or nanotube C causes magneto-thermoelectric current (MTEC) arising in a narrow metal stripe N which is crossing C in perpendicular direction. The MTEC arises from splitting the heat flow under influence an external d.c. magnetic field ${\bf B}=\{0,0,B_z\}$ which is applied to the C/N crossing (we call it the C/N-knot which is sketched in Fig.~\ref{fig1}). 
The heat flow emerges by injecting of non-equilibrium (NE) excitations from the external heater H into the C$^{\prime \prime}$ section of C  as shown in Fig.~\ref{fig1}a. The effective local temperature $T^{*}$ in C$^{\prime  \prime}$ is considerably higher than the equilibrium temperature $T_{\rm C}$ of the C ends. The NE injection causes a finite thermal current $Q = \Lambda \left( T^{*}-T_{\rm C}\right) $ along C where the heat conductance $\Lambda =\Lambda _{\mathrm{ph}}+\Lambda _{\mathrm{e}}+\Lambda _{\mathrm{h}}$ consists of the phonon ($\Lambda _{\mathrm{ph}}$), electron ($\Lambda _{\mathrm{e}}$), and hole ($\Lambda _{\mathrm{h}}$) components.~\cite{Pop,Cahill}  If number the NE electrons and holes which are excited in C is equal to each other (like, e.g., in undoped graphene or carbon nanotubes), the electric charge transferred along C is zero, i.e., $I_{\parallel} \equiv 0$. Another scenario might take place if the effective electron temperature $T^*$ is lower than the exciton binding energy $E_g$ whereas the NE charge carrier density $n^*$ is high. Then the electrons and holes could form the excitons which are also contributing to the heat transfer in C. In this Letter, however, we mostly consider the case $T^* > E_g$. Because in C (likewise in carbon nanotubes or in graphene) $\Lambda _{\mathrm{ph}}>>\max\{\Lambda _{\mathrm{e}},\Lambda _{\mathrm{h}}\}$, %
most of the heat along C is carried by phonons.~\cite{Pop,Cahill} The phonons also drag additional electrons and holes along C at expense of the phonon-electron collisions.~\cite{Bailyn,Bastard} 

Here we find that the Lorentz force acting in vicinity the C/N-knot splits the heat stream by pulling the electric charge carriers from C to N while the phonons are propagating further ahead. It results in a substantial electric current $I_{\perp} \neq 0$ which is formed in the lateral \^y-direction along N,  despite  $I_{\parallel} \equiv 0$. The non-equilibrium (NE) electron and hole excitations are created inside C by two mechanisms. (\textit{i})~Phonon drag the electrons and holes along C, and (\textit{ii})~Non-equilibrum thermal injection from the heat source H into C (see Fig.~\ref{fig1}a). The heat flow splitting in the C/N-knot becomes possible due to the following. We assume that the C$^{\prime }$/N- interface transparency $\mathcal{T}_{e,h}^{\rm CN}$ for electrons and holes transmitted from C$^{\prime }$ to N is high ($\mathcal{T}_{e,h}^{\rm CN}\lesssim 1$). Simultaneously, the same C$^{\prime }$/N-interface transparency $\mathcal{T}_{ph}^{\rm CN}$ is very low for phonons (i.e., $\mathcal{T}_{ph}^{\rm CN}<<1$) though their propagation inside C is  ballistic. That happens because redirecting the phonons from C to N requires a significant change of the phonon$^\prime $s momentum $\Delta {\bf q} = \hat {\bf y} q_{\rm N}-\hat {\bf x} q_{\rm C}$ where $q_{\rm N}$ and $q_{\rm C}$ are the phonon momentum components along N and C respectively.~\cite{Pop,Cahill} Acoustic properties of C and N are very different as well: Not only they have distinct phonon spectra but the geometries of C and N might differ from each other. For instance, C could be a cylinder shaped nanotube whereas the metal stripe has a bar shape. Therefore for the sake of simplicity we disregard the phonon transmission from C to N by setting $\mathcal{T}_{ph} \simeq 0$. Thus we assume that phonons merely pass the knot and propagate along C further ahead. On the contrary, we will see that the Lorentz force  easily redirects the non-interacting electrons and holes from C into the perpendicular metal stripe N. As it follows from further calculations, the redirection happens because the Lorentz force pulls electrons to the right (blue arrow in main Fig.~\ref{fig1}c) whereas it pushes the holes to the left (red arrow in Fig.~\ref{fig1}c). In this way, the heat flow (yellow arrow in Figs.~\ref{fig1}a,c) along C induces the electric current  $I_{\perp }$ (magenta arrow) in N  (brown).  
\begin{figure}
\includegraphics[width=75 mm]{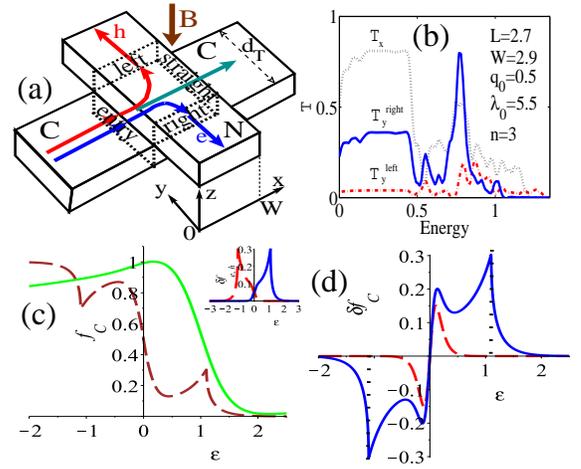}
\caption{Color online. (a)~A cross-stripe geometry used in addition to the 
nanotube/stripe cross geometry shown in Fig.~1c for computing the transmission probabilities through the C/N-knot. (b)~The transmission coefficients for an electron incoming from C into the C$^{\prime }$-section along the \^{x}-direction, $T_{x}$ (doted black curve), turning to the right, $T_{y}^{\rm right }$ (solid blue curve), and turning to the left, $T_{y}^{\rm left }$ (dashed red curve) computed for geometry (a) by solving non-linear boundary conditions. (c)~The non-equilibrium (NE) electron distribution function $f_C (\varepsilon )$ in the C section. Brown curve is  $f_C (\varepsilon )$ which is altered only by the NE phonon drag whereas green curve is  $f_C (\varepsilon )$ altered only by thermal injection the electron and holes (see parameters in text). Inset shows the phonon drag - induced NE deviation the distribution function $\delta f_{\rm e,h}$ of electrons (blue curve) and holes (red curve) in the C section for the same parameters. (d)~Driving factor for the lateral magnetoelectric current at $U_0 = 0$. Red dashed curve corresponds to influence of the NE thermal injection only while blue solid curve shows the combined influence of both, the NE thermal injection and the phonon drag.} 
\label{fig2} 
\end{figure}

We compute the electric current $I_{\perp }$ induced in the C/N-knot under influence the d.c. magnetic field ${\bf B} = (0,0,B_z)$. Let us assume that an electron which emerges from the left open C section then penetrates into the C$^\prime $ section under metal N (see Fig.~\ref{fig2}a). At the next stage there are actually three further possible distinct trajectories for an electron to pass through the knot. (a)~The electron is then transmitted into the right wing of N (blue arrow in Fig.~\ref{fig2}a); (b)~It can also proceed straightforward through the knot from C$^\prime $ to C on the opposite right hand side (green arrow in Fig.~\ref{fig2}a); (c)~Instead of being transmitted from C$^\prime $ to the right N-wing, an electron gets into the `wrong' left N-wing (red arrow in the same figure). We emphasize that only the $e$- and $h$-deflected trajectories (see correspondly the blue and red lines in Fig.~\ref{fig1}c) which energy diagram is shown in Fig.~\ref{fig1}e contribute into the electric current generated along N. They correspond to an asymmetric C/C$^{\prime }$/N-junction sketched in Fig.~\ref{fig1}e. On the contrary, the straightforward tragectory shown by the green line in Fig.~\ref{fig1}c and which is related to a symmetric C/C$^{\prime }$/C junction (see energy diagram in Fig.~\ref{fig1}d) gives no contribution into the electric current generated along N. The C/C$^{\prime }$ and C$^{\prime }$/N interfaces might additionally contain potential barriers I originating from atomic impurities (shown as brown in Fig.~\ref{fig1}a,b,d,e), imperfections, and/or difference the workfunctions of N and C. The lateral electric current generated in N is calculated using the Landauer-B\"uttiker formula%
\begin{eqnarray}
I_{\perp} = I_{\rm left}-I_{\rm rigth}=\frac{2e}{h}\int_{0}^{\infty } d\varepsilon M_{\min }\left( \varepsilon \right) \nonumber \\
\times 2 \left ({\cal T}_{y}^{\rm left} \left( \varepsilon
\right) - {\cal T}_{y}^{\rm right}\left( \varepsilon
\right) \right) \left( f_{C}\left( \varepsilon -\mu \right) -f_{N}\left(
\varepsilon \right) \right)  \label{current}
\end{eqnarray}%
where $M_{\min }\left( \varepsilon \right) $ is the minimum number of
quantum channels per energy interval, ${\cal T}_{y}^{\rm left (right)}\left( \varepsilon \right) $ is the C/C$^{\prime }$/N knot transparency for an electron to penetrate from section C through section C$^{\prime }$ into the left (or right) wing of N, $\mu \simeq U_0 $ is the electrochemical potential shift inside C.  A finite electric potential $U_{0}$ arises, e.g., due to coupling between the electron orbitals in N and C$^{\prime }$. An extra factor 2 in Eq.~(\ref{current}) comes from the hole contribution into the lateral electric current $I_{\perp}$. 

The lateral electric current $I_{\perp }$ induced
by the phonon drag and the thermal injection~\cite{TEG-graph} depends on electron and phonon distribution functions $f_{\rm C}$ and $N_{\rm C}$  inside the C$^{\prime }$ section. The kinetic equations are%
\begin{eqnarray}
\dot{f}_{\rm C} = \mathcal{L}_{th} + \mathcal{L}_{ep} \label{kin-el} \\
\dot{N}_{\rm C} = \mathcal{P}_{th} + \mathcal{P}_{pe}  
\label{kin-eq}
\end{eqnarray}%
where  $%
\mathcal{L}_{th}$ and $\mathcal{P}_{th} $ describe thermal injection of the non-equilibrium electrons, holes, and respectively phonons from the heat source H into the C$^{\prime \prime}$ section as shown in Fig.~\ref{fig1}a, $\mathcal{L}_{ep}$ is the electron-phonon collision integral,  whereas $\mathcal{P}_{pe}$ is the phonon-electron collision integral. In Eqs.~(\ref{kin-el}), (\ref{kin-eq}) we have disregarded other processes having a lesser importance. Phonon drag is essential in semiconducting nanotubes (stripes) whereas it is negligible in metallic nanotubes (stripes). The drag occurs~\cite{Tsao,Bailyn} when  NE phonons with frequency $\Omega _{0}$ propagating between the hot (C$^{\prime \prime }$) and cold (C$^{\prime }$) spots of the same C generate electrons and/or holes moving in the same direction during the phonon-electron collisions. In Eq.~(\ref{current}), the electron distribution functions $f_{C} = f_{C}^{(0)} (T_{\rm C}^{\prime \prime}) + \delta f^{\rm th} + \delta f^{\rm pd}$ significantly deviates from its equilibrium value $f^{(0)} (T_{\rm C}) =1/(\exp{(\varepsilon /T_{\rm C}}) +1) $ due to the thermal injection from H to C$^{\prime \prime }$ by $\delta f^{\rm th}$ and due to the phonon drag generation by $\delta f^{\rm pd}$. The NE deviations $\delta f^{\rm th}$ and $\delta f^{\rm pd}$ are computed using Eqs.~(\ref{kin-el}), (\ref{kin-eq}). The electron distribution function $f_{\rm N}$ taken in the left and right wings of the N-stripe $f_{\rm N}^{\rm left/right}$ do not much differ from each other (i.e., $f_{N}^{\rm left} \simeq f_{N}^{\rm right}$). Therefore we merely regard them as equilibrium by setting $f_{\rm N} \simeq f^{(0)}(T_{\rm C}^{\prime })$ and characterize them by temperature $T_{\rm C}^{\prime }$. Besides we neglect by the jump of electric potential along N coming across the C/N-knot in the \^y-direction. Eq.~(\ref{kin-el}) with $\mathcal{L}_{th}$ had been addressed in Ref.~\cite{TEG-graph} whereas the phonon drag had been considered in Refs.~\cite{Bailyn,Tsao,Shafr-TEG}. On one hand, a small deviation $\delta f^{\mathrm{pd}}$ is obtained from Eq.~(\ref{kin-el}) as
\begin{equation}
\delta f^{\rm pd}\simeq  \pi \tau _{ep} g^{2}\sum_{i=\pm }\left\vert m\left(q_{i}\right) \right\vert ^{2}\delta N(sq_{\pm}) [f_{i} + 
f^{\left( 0\right)}-2f^{\left( 0\right) }f_{i}] 
\label{dfph}
\end{equation}
where $\tau _{\rm ep }$ is the electron-phonon collision time, $f^{\left( 0\right) }\left( \varepsilon \right) =1/\left( \exp \left(
\varepsilon /T\right) +1\right) $,  $f_{\pm}\left(
\varepsilon \right) =1/\left( \exp \left( \sqrt{\varepsilon
^{2}+v^{2}q_{\pm}^{2}+2q_{\pm}v\sqrt{\varepsilon ^{2}-v^{2}q_{\nu }^{2}}}%
/T\right) +1\right) $, $q_{\pm }=2\left( k\pm \left( s/v\right) 
\sqrt{k^{2}+q_{\nu }^{2}}\right) /\left( \left( s/v\right) ^{2}-1\right) $, $s$ is the speed of sound, and $v$ is the Fermi velocity in C.
On the other hand, the thermal injection changes $f_{\rm C}$ as 
\begin{equation}
\delta f^{\mathrm{th}} = \frac{\tau _{\rm ep } \Gamma _{\mathrm{HC}}}{1 + \tau _{\rm ep } \Gamma _{\mathrm{HC}}}(f_{\rm C^{\prime \prime}}-f^{(0)})
\label{dfth}
\end{equation}%
where $\Gamma _{\mathrm{HC}}$ is the electron tunneling rate between H and C, $f_{\rm C^{\prime \prime}} = 1/(\exp{(\varepsilon /T_{\rm C^{\prime \prime }})}+1)$. 
The NE deviation $\delta N = N_{\rm C}-N^{(0)}$ from the Bose-Einstein function $N^{(0)}$ inside C entering (\ref{dfph}) is found from Eq.~(\ref{kin-eq}) in the form
\begin{equation}
\delta N\left( \omega \right) = N_{0}\exp \left( -\left( \omega -\Omega
_{0}\right) ^{2}/\sigma _{\rm ph}\right)  \label{del-N}
\end{equation}%
where the parameters $N_{0}$, $\Omega _{0}$, and $\sigma _{ph}$ are the amplitude, frequency, and width of the NE phonon beam propagating along C from C$^{\prime \prime }$ to C$^{\prime }$. They are determined by boundary conditions, computed microscopically, and/or extracted from the experiment. Intensity $N_0$ the phonon beam (\ref{del-N}) inside C is determined by power  the heat source H, and by the H/C-interface transparency ${\cal T}_{\rm ph}^{\rm HC}$. The beam width $\sigma_{\rm ph}$ depends on the geometry of H.

\textit{Transmission probability} through the C/N-knot is computed using two different geometries depicted in Figs.~\ref{fig1}c and \ref{fig2}a. As a trial wavefunction $\Psi_{\rm C^\prime } \left( x,y\right)$ for cross-stripe geometry in Fig.~\ref{fig2}a we use a two-dimensional electron wave inside the rectangular box C$^\prime $ under the metal N. We assume that an electron with the \^x-component momentum $k$ enters the C/N-knot on the left side from C to C$^\prime $ at $x=0$ and $0<y<d_{\rm T}$. The electron exits the C/N-knot to N either on right ($y=0$, $0<x<W$), on left ($y = d_{\rm T}$, $0<x<W$), or to C straight ($x=W$, $0<y<d_{\rm T}$) where $W$ is the metal stripe width, and $d_{\rm T}$ is the C diameter. While dwelling in C$^\prime $ under metal N, the electrons (holes) acquire a transversal momentum component $q = \pm \left( \left( e/c\right) B_z y + t^{2}g/v\right) $ where the former term is due to magnetic field whereas the last term is caused by the C/N tunneling coupling with energy $t^{2}g$. Here $t$ is the tunneling matrix element, $g$ is the electron density of states in the metal stripe N, $v$ is the Fermi velocity in C, and $B_z$ is the magnetic induction directed along z-axis. For the rectangular geometry of Fig.~\ref{fig2}a the \textit{trial wavefunction} in C$^\prime $ under metal is constructed as%
\begin{equation}
\Psi_{\rm C^\prime } \left( x,y\right) =\left( \text{$\alpha $}_{x}e^{ikx}+\text{$\beta $}%
_{x}e^{-ikx}\right) \left( \text{$\alpha $}_{y}u_{n}\left( \zeta_{+}\right) +%
\text{$\beta $}_{y}u_{n}\left( \zeta_{-}\right) \right)
\label{psi}
\end{equation}%
where we have introduced auxiliary functions $u_{n}\left( \zeta \right) =\exp \left( -\zeta^{2}/2)\right) H_{n}\left(
\zeta\right) $, $H_n(\zeta )$ is the Hermite polynomial, $\zeta_{\pm }=\zeta\left( y,\pm k\right) =\sqrt{m\omega _{c0}/\hbar }\left(y\mp \beta ^{2}k\right) = q_{0}y\mp \lambda _{c0}k$. 
Here $q_{0}=\sqrt{m\omega _{c0}/\hbar }$, $\lambda _{c0}=(\omega _{c}/\omega _{c0}) ^{2}\lambda ^{2}q_{0}$, $\lambda = \sqrt{\hbar /eB}$, $\omega_c = eB/m_e$ is the cyclotron frequency, $m_e$ is the electron mass. The energy dependence of $k$ then is $k (\varepsilon) = (\omega _{c0}/\omega _{0})(\sqrt{2m\Delta _{n}}/\hbar )\sqrt{(\varepsilon -\Delta _{n})/\Delta _{n}}$
where $\Delta _{n}=E_{s}+\hbar \omega _{c0}\left( n+1/2\right) $ and 
$\omega _{c0}=\sqrt{\omega _{c}^{2}+\omega _{0}^{2}}$, $\omega _{0}$ is the characteristic of steepness the confinement potential. In the above formulas we disregard the chirality effects \cite{Ando} since the H/C$^{\prime \prime }$ and C$^{\prime }$/N transmissions are non-chiral. The trial wavefunctions in C and N outside the C/N-knot region are taken as mere plane waves. The boundary conditions (BC) yield the system of 8 non-linear transcendental equations for 8 coefficients entering the trial wavefunctions. The BC equations had been solved numerically which allows computing of the transmission probabilities ${\cal T}_{y}^{\rm left (right) }$ and ${\cal T}_{x}$ entering Eq.~(\ref{current}). 

Using the above non-linear boundary conditions for the cross-stripe setup (Fig.~\ref{fig2}a) to computing the transmission probabilities through the C/N-knot we have studied transport properties the C/N-knot. In Fig.~\ref{fig2}a we show transmission coefficients for an electron incoming from C into the C$^{\prime }$-section along the \^{x}-direction, $T_{x}$ (doted black curve), turning to the right, $T_{y}^{\rm right }$ (solid blue curve), and turning to the left, $T_{y}^{\rm left }$ (dashed red curve) obtained as solutions of the non-linear boundary conditions. Similar results had also been obtained for the nanotube-stripe (Fig.~\ref{fig1}c) geometry. We show the numeric results in Fig.~\ref{fig2}b where we also indicate parameters of the C/N-knot and the external magnetic field $B_z$. One can see that the transmission probabilities $T_{y}^{\rm left (right) }$ and ${\cal T}_{x}$ are both experiencing strong resonances which suggest strong deflecting the electrons and holes from C to N in the lateral \^y-direction at certain values the electron energy $\varepsilon $ and magnetic field $B_z$. To emphasize the difference between the two NE mechanisms (i.e., phonon drag and thermal injection) we set either $\delta f^{\rm pd} \equiv 0$ while $\delta f^{\rm th} \neq 0$ or otherwise $\delta f^{\rm th} \equiv 0$ while $\delta f^{\rm pd} \neq 0$. In Fig.~\ref{fig2}c we plot the non-equilibrium (NE) electron distribution function $f_C (\varepsilon )$ in the C section obtained as a self-consistent numeric solution of Eqs.~(\ref{kin-el}), (\ref{kin-eq}). Brown curve is  $f_{\rm C} (\varepsilon )$ altered only by the NE phonon drag whereas green curve is  $f_{\rm C} (\varepsilon )$ altered only by thermal injection the electron and holes at $\Gamma_{\rm HC} = 0.3$. We have used the following parameters   $\tau_{\rm ep} = 10^{-12}$~s, $\sigma _{0}=0.2$, $N_{0}=1$, $\Delta_{v} = 1$, $T_{\rm C}^\prime = 0.1$ is the temperature in C$^\prime $-section, $s = 0.02$ is the sound velocity, $v = 1$. In Fig.~\ref{fig2}b and below, for the sake of convenience, the characteristics with energy and temperature dimensions are expressed in units the semiconducting gap  $\Delta_{v}$ in C (we set $\hbar = 1$). Parameters with the wavenumber dimensions (like $k$, $q_0$, $q_c$, etc.) are expressed in units  $\Delta_{v}/v$ whereas the length parameters (like $L$, $W$, $\lambda $, $\lambda_{\rm c0}$, etc.) are expressed in the units of $v/\Delta_{v}$. Then, e.g., for a carbon nanotube (CNT) with diameter $d_{\rm T} = 2.5$~nm one gets $\Delta_v = 2\hbar v/d_{\rm T} = 0.4$~eV, $\Delta_{v}/v = 7.5 \times 10^8$~m$^{-1}$ and $v/\Delta_{v} = 1.3$~nm where for CNT (and graphene) we also use that   $v=8.1 \times 10^5$~m/s and $s=2.1 \times 10^4$~m/s. Inset in Fig.~\ref{fig2}c shows the phonon drag - induced NE deviation the distribution function $\delta f_{\rm e,h}$ of electrons (blue curve) and holes (red curve) in the C section for the same parameters as listed above. Driving factor $\delta f_{\rm C} = f_{\rm C}-f_{\rm N}=\delta f^{\rm pd}+\delta f^{\rm th}$ for the lateral magnetoelectric current $I_{\perp }$ at $U_0 = 0$ is plotted in Fig.~\ref{fig2}d. Red dashed curve corresponds to influence of the NE thermal injection only (we set for the moment $\delta f^{\rm pd} \equiv 0$) while blue solid curve shows the combined influence of both, the NE thermal injection and the phonon drag.

\begin{figure}
\includegraphics[width=95 mm]{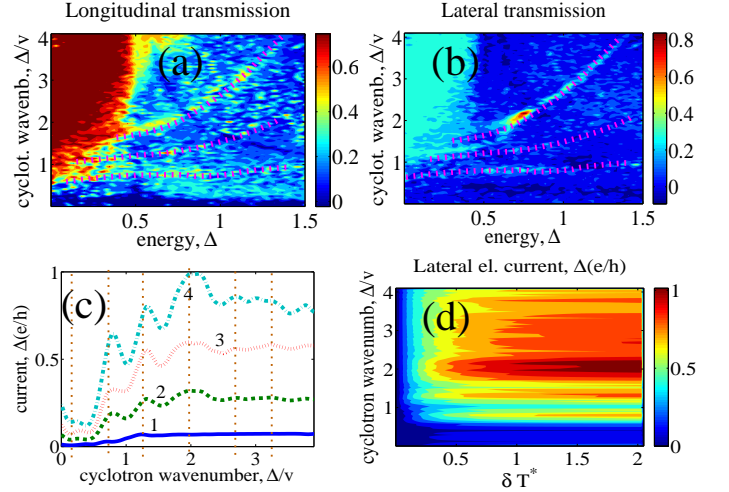}
\caption{Color online. (a)~Longitudinal transmission probability through the C/N-knot versus the electron energy and the cyclotron wavenumber $q_c = \sqrt{eB_z/\hbar }$ (see parameters in text).(b)~ Lateral transmission probability computed for the same C/N-knot parameters. (c)~Lateral magneto-thermoelectric current (MTEC) $I_{\perp}$ induced along N versus the cyclotron wavenumber $q_c$ for different temperature deviations $\delta T^*$ (see text) in C. (d)~Corresponding contour plot of steady state MTEC for the same C/N-knot parameters.} 
\label{fig3} 
\end{figure}
In Fig.~\ref{fig3} we show transmission probabilities and MTEC versus magnetic field $B_z$ and temperature deviation $\delta T^* = T^*-T_{\rm C}$ where $T_{\rm C}$ is the steady state temperature of C. The effective electron temperature $T^*$ which meters the NE effect is introduced as $T^* = \int_0^\infty d\varepsilon \cdot \varepsilon M_{\rm min} \delta f_{\rm C}$. In Fig.~\ref{fig3}a we plot the longitudinal transmission probability through the C/N-knot versus the electron energy $\varepsilon $ and the cyclotron wavenumber $q_c = \sqrt{eB_z/\hbar }$. The C/N-knot length and width are $L = 2.7$ and $W = 2.9$ respectively, $q_0 = 0.5$, $\lambda_0 = 5.5$, $U_0 = 0.5$. Computed lateral transmission probability is shown in Fig.~\ref{fig3}b for the same C/N-knot parameters. In Fig.~\ref{fig3}c we plot the lateral magneto-thermoelectric current (MTEC) $I_{\perp}$  in units $\Delta (e/h)$ induced along N versus the cyclotron wavenumber $q_c$  (in units $\Delta/v$) for different effective temperature deviations  $\delta T^* = 0.08, 0.8, 1.4$, and 2 in units $\Delta $ for curves 1-4 respectively) along the \^x-axis in C. As an illustration, we also show the corresponding contour plot of MTEC for the same C/N-knot parameters.
Maximums the heat-generated electric current in N correspond to geometrical resonances when different Landau orbits match the C/N-knot dimensions. From Fig.~\ref{fig3}d one can also see that $I_{\perp}$ is roughly proportional to the temperature deviation $\delta T^*$ in C. The magnetic field induction $B_z$ and the temperature deviation $\delta T^*$ both facilitate increasing the electric current $I_{\perp}$ (see, i.e., Figs.~\ref{fig3}c,d) at the expense of non-equilibrium phonon drag and thermal injection. 

The obtained results suggest that after penetrating from C$^\prime $ to N under influence the Lorentz force (see Fig.~\ref{fig1}b), the electrons (e) and holes (h) propagate inside N in opposite directions. Since the electrons and holes also have opposite electric charges, they create a substantial electric current $I_{\mathrm{e}}=I_{\mathrm{h}}=I\neq 0$ flowing along N. Magnitude the thermally induced electric current in N is proportional to the heat flow in C and it also depends on the d.c. magnetic field ${\bf B}$. A finite field ${\bf B} \neq 0$ splits the heat flow and separates the electrons and holes from each other as is shown in Fig.~\ref{fig1}a.

The consideration above neglects the role of electron-electron and electron-hole interaction in thermal transport. However many-body effects might be essential in a system where the NE electron density $n^* = \int_0^\infty d\varepsilon \cdot M_{\rm min} \delta f_{\rm C}$ is high. One application of the suggested here heat stream splitting is to probing the electron-electron interaction and determining the exciton binding energy $E_g$. For low temperatures and small $\delta T^*$ when $E_g > T_{\rm C}$, the excitons transfer the heat from C$^{\prime \prime}$ to C$^{\prime }$. 
Since the excitons have no electric charge, there is no splitting of heat flow in the C/N-knot, and no lateral electric current ($I_{\perp } \equiv 0$) is induced in N. However $I_{\perp } $ turns being finite as soon as the local temperature $T_{\rm C}^{\prime} \leq T^*$ in the C/N-knot vicinity exceeds the threshold value $T_{\rm C}^{\rm thr} $ which corresponds to the binding energy $E_g$, i.e., $T_{\rm C}^{\rm thr} \simeq E_g < T_{\rm C}^{\prime } \leq T^*$ and no excitons are formed in C any more. Thus presence of  the threshold temperature $T_{\rm C}^{\rm thr}$ which turns $I_{\perp } $ on indicates creating of excitons inside C with the binding energy  $E_g \simeq T_{\rm C}^{\rm thr}$. The considered here approach might provide an efficient thermoelectric solution which exploits splitting the thermal current components by so-called C/N-knot. Suggested generation of electric current potentially can be used in thermoelectric energy generators and coolers.~\cite{TEG-book,P-Kim,Dressel,Shafr-TEG} 

I wish to thank P.~Kim and V.~Chandrasekhar for fruitful discussions. This work had been supported by the AFOSR grant FA9550-11-1-0311.

\end{document}